\documentclass[aps,pra,twocolumn,groupedaddress,showpacs]{revtex4}
\usepackage{graphicx}
\usepackage{dcolumn}
\usepackage{bm}
\begin{document}
\title{
Quantum stream cipher by Yuen 2000 protocol: 
Design and experiment by intensity modulation scheme
}
\author{Osamu Hirota}
\altaffiliation[Also at ]{
21st century COE program, Chuo University
}
\email{
hirota@lab.tamagawa.ac.jp
}
\affiliation{
Research Center for Quantum Information Science, 
Tamagawa University\\
6-1-1, Tamagawa-gakuen, Machida, Tokyo, 194-8610, JAPAN
}
\author{Masaru Fuse}%
\email{
fuse.masaru@jp.panasonic.com
}
\affiliation{
Panasonic (Matsushita Electric Industrial Co., Ltd.), Osaka, JAPAN
}
\author{Kentaro Kato}
\email{
kkatop@ieee.org
}
\affiliation{
21st century COE program, Chuo University, Tokyo, JAPAN
}
\author{Masaki Sohma}
\email{
sohma@eng.tamagawa.ac.jp
}
\affiliation{
Department of Information Science, 
Tamagawa University, Tokyo, JAPAN
}
\date{\today}
\begin{abstract}
This paper shall investigate Yuen protocol, so called Y-00, which can realize a randomized stream cipher with high bit rate(Gbps) for long distance(several hundreds km). The randomized stream cipher with randomization by quantum noise based on Y-00 is  called quantum stream cipher in this paper, and it may have  security against  known plaintext attacks which has  no analog with any conventional symmetric key ciphers. 
We  present a simple cryptanalysis based on an attacker's heterodyne measurement and the quantum unambiguous measurement to make clear the strength of Y-00 in  real communication. In addition, we give a design for the implementation of an intensity modulation scheme and report the experimental demonstration of 1 Gbps quantum stream cipher through 20 km long transmission line.
\end{abstract}
\pacs{03.67.Dd, 42.50.Lc}
\keywords{Quantum cryptography, Data encryption}

\maketitle

\section{Introduction}
It is very difficult to devise  encryption schemes with "provable security" in the conventional cryptography. So far we have two schemes for encryption with provable security. 
 One of methods is one time pads supported by the quantum key distribution. The other is a kind of randomized stream cipher. Recently, most of the efforts to realize the encryption with provable security have been devoted to the quantum key distribution invented by C.Bennett and G.Brassard in 1984[1]. 
We emphasize the greatness of this achievement which opened a new scientific realm.
However, there is a big gap  between the  experimental realization and the real communication network requirement. In addition, unfortunately one may say there is no practical unconditionally secure protocol that has ever even been theoretically proposed. Furthermore, any kind of quantum repeater cannot guarantee high key rate. The key rate decreases exponentially with respect to the distance[2]. This might be equivalent to no repeater scheme. Even if the quantum repeater with a quantum media transform is employed[3], the improvement is so little, because  no perfect quantum efficiency of media transformation exists. 
There is no means of improving such a poor performance.
So we would like to  point out that the key generation is  very important, but  is very narrow minded to define quantum cryptography only by BB-84 and  similar principle[4]. 
Thus, it is preferable to investigate a quantum symmetric key cipher with information theoretic security based on quantum/optical communication. In 2000, Yuen announced that his new protocol so called Yuen protocol (Y-00) may provide a randomized stream cipher with information theoretic security by randomization based on quantum noise and additional mathematical schemes[5, 6]. This scheme is called quantum stream cipher, or $\alpha \eta$ scheme by the Northwestern University group.
In  conventional cryptography, there is no known complexity-based proof at all  on any scheme under  known plaintext attacks on key.  It is an interesting subject to show, by the concrete scheme, that the quantum stream cipher by Y-00 has a potential of the information theoretic security against  known plaintext attacks.

In the proceedings of SPIE in Denver[7] and of Quantum Informatics in Moscow[8], we gave a framework of the concrete security analysis for quantum stream cipher by Y-00. 
In this paper, we shall show how to apply the results of these papers [7,8] to security analysis, and a design and experimental demonstration of the quantum stream cipher by an intensity modulation. However, the general proof of the security of the quantum stream cipher by Y-00 still remains. The direction of the proof has been suggested by Yuen in his paper[6].

\section{Information theoretically secure 
stream cipher}
First, we will denote the definition of the information theoretic security. Let us assume that the eavesdropper Eve has the following abilities.
\begin{itemize}
\item[\rm(i)]  She has the computer with unlimited computation power
\item[\rm(ii)] She has the unlimited memory capacity
\end{itemize}
When Eve cannot decode  plaintext or key even if she has the above computational power, the cryptography is information theoretically secure. Many works on  protocol with information theoretic security  have been already published in journals of information theory and cryptography. In the following, we will introduce some examples.

\subsection{One time pad}
The definition of  perfect secrecy, that is information theoretic security against any kind of the criteria, is $H(X|Y)=H(X)$, where $X$ and $Y$ are plaintext and ciphertext, respectively. 
It  means that the plaintext $X$ and the ciphertext $Y$ as a function of $X$ are statistically independent. 
In order to realize such a perfect secrecy, the condition: $H(X)\le H(K)$ is required[9], whenever Eve has an access to precisely the same information as the legitimate users. This situation, in which Eve and Bob can get the same ciphertext, is reasonable in the conventional communication network.
In this situation, one of methods to realize the perfect secrecy is one time pad or Vernum cipher which is a kind of stream cipher. However, as mentioned above, it requires the secret key which is at least as long as the plaintext message. 
If the infinite key for one time pad can be sent through secure communication, then the one time pad makes sense in the real communications. So far, many researchers in quantum information science have proposed protocols in order to realize a secure key distribution which are guaranteed by quantum effect in communication process. BB-84, E-91, and B-92 are typical examples of such protocols. 

\subsection{Randomized stream cipher}
In the conventional cryptosystem, the stream cipher is implemented by a pseudo random number generator with short secret key and XOR operation with plaintext data bit.
For a symmetric key cipher  as a direct data encryption, the main criteria of the security are given as follows.
\begin{itemize}
\item[\rm(i)]  Ciphertext-only attack(CTA) on data and on key: To get plaintext or key, Eve can know only the ciphertext from her measurement.
\item[\rm(ii)] Known/chosen plaintext attack(KTA): To get key,  Eve  can know  non-uniform statistics for some plaintexts and corresponding ciphertexts, or insert chosen plaintext data into the modulation system( for example, inserts all 0 sequence as plaintext in some periods). 
Then Eve tries to determine the key from input-output. Using the key, Eve can determine the remained data from the ciphertexts.
\item[\rm(iii)] Repetition attack: Since the secret key is fixed, it has a period. Eve can apply CTA and KTA over many periods when the key is reused.
\end{itemize}

We can summarize the performance of the conventional stream cipher by the unicity distance defined by Shannon[9].
For the ciphertext only attack on key, the unicity distance is as follows:
\begin{equation}
n_u=\min\{n:H(K|Y^n)=0\}
\end{equation}
For the known plaintext attack, it can be modified into 
\begin{equation}
n_{Gu}=\min\{n:H(K|Y^n, X^n)=0\}
\end{equation}
As an example, the unicity distance is sometimes given as follows:
\begin{equation}
n_u \sim \frac{H(X)}{D}
\end{equation}
where $D$ is the redundancy of the plaintext sequence.
When the statistics of plaintext is uniform, the unicity distance becomes infinite. 
Shannon called ciphers with $n_u=\infty$ "ideal cipher". 
This is the  information theoretic security against  ciphertext only attacks on data. 
However, the unicity distance of many conventional stream ciphers against  known plaintext attacks is finite and some times it is 
\begin{equation}
u_{Gu}\sim H(K)
\end{equation}
For example, let us use a linear feed back shift register(LFSR) as a pseudo random number generator. Eve can know the secret key when she get $2|K|$ bits (key and shift parameter uncertainty) as the running key from a pseudo random generator. 
Thus the stream cipher is surely broken by a brute force attack only for known plaintext attacks, but not for ciphertext only attacks and not for known non-uniform plaintext statistics attacks when the length of known plaintext is smaller than the unicity distance.
Although there are several proposals which have better performance than that of Eq(4), no one succeed to show the lower bound. 
Again, the symmetric key ciphers  are in principle insecure. This derives from the fact that the security is given only by key uncertainty. 

There is another theoretical issue in the discussions on the information theoretically secure stream cipher. It is called "randomized stream cipher", which was long known, even to Gauss. The randomized stream cipher means that ciphertexts are randomized. 
In the modern cryptography, they are designed based on information theoretic approach, and are discussed by  C.Schnorr, Diffie, Maurer, and Cachin[10]. 
Maurer devised a randomized stream cipher for which one can prove that Eve obtains no information in Shannon's sense about the plaintext with probability close 1. But his protocol works under the assumption that memory capacity of Eve is limited[11]. 
This approach provides an information theoretic notion of security under a computational restriction. However, unfortunately, it is difficult to implement the practical system with high speed processing.
Thus, it is very difficult to realize the information theoretically secure symmetric key cipher based only on mathematical algorithm.
Yuen, however, points out that it may be possible when one employs randomization by quantum noise. In the following sections, we will show the concrete scheme and the performance of Yuen protocol so called Y-00.

\section{Quantum communication  for quantum cryptography}
In any quantum cryptography such as BB-84, B-92, E-91, and Y-00,  information is classical bit sequence. That is,  information is the true random bits for the key distribution, and the plaintext bits for the direct data encryption.
The essential assumption in quantum communication for classical information is that quantum states are known to the legitimate users. So classical bits are mapped into a set of known quantum states. 
They are transmitted passing through a completely positive map (cp-map), and  discriminated by quantum measurement process described by positive operator valued measure(POVM). Then, a receiver gets classical bits as information by measurements. This model is called Helstrom/Holevo/Yuen formalism for the quantum communication[12,13,14].
Let us give a brief introduction.
Source and output in the quantum communication model are described by a density operator for ensemble of quantum states which conveys classical information as follows:
\begin{equation}
\rho_{Tin} =\sum p_i \rho_i,\quad \rho_{Tout}= \sum p_i \epsilon(\rho_i),
\end{equation}
where $i$ is an index corresponding to symbol as classical information, and $\epsilon$ is a cp-map.
The discrimination among quantum states at the output of the channel is described by POVM.
\begin{equation}
\Pi_j \ge 0, \quad \sum \Pi_j =I,
\end{equation}
where $I$ is an unit operator.
Then a conditional probability for each trial of the measurement is given by 
\begin{equation}
P(j|i)=Tr \epsilon(\rho_i) \Pi_j.
\end{equation}
The minimization problem of the average error probability based on the above equation is called quantum detection theory, which is a fundamental formalism in quantum information science.
\begin{equation}
P_e =\min_{\Pi} \{1-\sum p_i Tr\epsilon(\rho_i )\Pi_i\}
\end{equation}
The complete theory has been given by 
Helstrom, Holevo, and Yuen. 
As a result, we have [12, 13, 14]\\
\\
{\bf Theorem 1}: 
{\it Signals with non-orthogonal states cannot be 
distinguished without error and  optimum lower bounds for error rate exist.}\\
\\
When the error  probability is 1/2, there is no way to distinguish them. The other important one is no cloning theorem clarified by
Wootters-Zurek, Yuen, and Buzek-Hillery [15] as follows.\\
\\
{\bf Theorem 2}: \\
{\it Non-orthogonal states cannot be cloned with perfect fidelity and with probability 1}\\
\\
The most important cp-map (communication channel) in the real world is energy loss channel with $20 \sim 100 dB$ loss. A selection of  input quantum states for the channel is one of interesting problems in quantum communications. But we have the following result[16].\\
\\
{\bf Theorem 3}: 
{\it The input state which keeps pure state passing through energy loss channel is only coherent state.}\\
\\
So we can understand that a desirable state is a coherent state.
The question is whether a coherent state is appropriate or not 
when we take into account two criteria: the efficiency and the security  as requirements to quantum communication for quantum cryptography. Y-00 protocol will verify that the communication by a coherent state can satisfy these two criteria.

\section{Yuen protocol:Y-00}
\subsection{Basic concept}
A symmetric key cipher is a scheme that the legitimate users, Alice and Bob, share a secret key. A block cipher and a part of stream ciphers belong to this category. However, they are in principle insecure, because the security is given only by  key uncertainty. The problem is whether we can realize the information theoretically secure cipher under a coherent state system with the finite secret key.
In the quantum communication model for quantum cryptography, we have to consider two channels of Alice-Bob, and of Alice-Eve.
Let us describe them by $\epsilon_{AB}$, $\epsilon_{AE}$.
In general, $\epsilon_{AE}$ is an ideal channel while $\epsilon_{AB}$ is a noisy channel. Even so, the basic performance of cryptography is to prevent the leak of secret information from channel of legitimate users.
In a physical cryptography like quantum cryptography, one may take a method to eliminate Eve's information obtained by her measurement from $\epsilon_{AE}$. In order to realize such a situation, one needs "advantage creation" under the ultimate physical law. It means that the disadvantage of Bob can be got rid of by some processing, while the performance of Eve, who has the unlimited power of computer and physical resources, is superior to that of Bob's in the original situation.

According to  quantum detection theory we have the following properties for the average error probability:
\begin{equation}
P_e(BP) < P_e(BM), \quad P_e(BP) < P_e(MP)
\end{equation}
where $BP$, $BM$, $MP$ mean binary pure state, binary mixed state, and $M$-ary pure state, respectively. The problem is  how to apply the above principle of the quantum detection theory to cryptography.
Yuen proposes a protocol which combines  a  shared secret  key  for the legitimate users and  specific quantum state modulation scheme. {\it This can be called initial shared key advantage} in a noisy channel. 
By this advantage, the legitimate users can establish  "the advantage creation" under  parameters  with the finite size in the protocol for noisy channel. 
As a result, one can see a basic principle to guarantee the security  as follows[6]:\\

{\bf Principle of security }: 
{\it The optimum quantum measurements with key and without key have different performance.}\\

An unknown key corresponds to the classical randomness. The security of the conventional symmetric key cipher is guaranteed by  this classical randomness. However, in Yuen protocol, a classical randomness is used to make a difference in the performances of quantum measurements. It means that if Eve does not know the key,  the quantum limitation of her measurement is enhanced by the classical randomness. As a result, Eve has to search for the data or the key based on her measurement results with the unavoidable error.
For an explanation of this principle, Yuen gives a simple example in the case without any design as follows[6]; Let us assume that the information source is  binary coherent states, and the ultimate error performances for a receiver with key or without key are
\begin{equation}
{P_e}^B \sim \exp(-4S) \quad {\rm vs} \quad {P_e}^E \sim \exp(-2S)
\end{equation}
where $S=<n>$ is signal photon per pulse.  
These are the error rates derived from the condition that the key is known to Eve after her measurement, or that she uses her measurement result to pin down the data for various different possible key values.
This fact gives us an advantage creation under the ultimate physical law, so it leads to the unconditionally secure key generation for the any key length of the initial key, and it also  gives a basis for  information theoretically secure direct data encryption.
The above example is not what we use as a practical quantum cryptography. It only shows  a principle. For the practical use, we need several additional contrivances. 
The essential problem is  how to extend the above principle towards the practical quantum cryptography.

The first idea was proposed as follows[17,18]. Let us prepare $M$ sets of two coherent states with a phase difference $\pi$. These are  the basis for the transmission of data.
We assume that Alice and Bob share a secret  key $K$. The key is stretched by a pseudo random number generator(PRNG). Let $K'$ be the pseudo random number from PRNG.
The random decimal number with $mod  M$ generated from the block: 
\begin{equation}
K'/\log_2 M \equiv \bar{K}'=(\bar{k_1}',\bar{k_2}',\dots)
\end{equation}
of pseudo random number is called the running key. A basis is randomly selected by the running key $\bar{K}'$. The data bit is transmitted by the selected basis.
The numbering of the basis set is $\{1,2,3.\dots \}$ from around $\theta=0$ to $\theta=\pi$ on the phase space.
As a result, the $M$-ary keying has $M$ different basis based on 2$M$ coherent states. So the data bit is mapped into one of 2$M$ coherent states randomly, but of course its modulation map has the definite relationship, which is opened. 
Bob knows the key and the running key, so his measurement is always the correct binary detection for signals with large signal distance. Since Eve does not know the key, she has to employ basically $M$-ary detection or other methods. But these are not better than Bob's measurement.

\subsection{Signal design}
Alice and Bob in Y-00 share a secret key $K$. The key is stretched by  a pseudo random number generator. The length of the running key is $|K'|$.
The data bit is modulated by $M$-ary keying driven by random decimal numbers generated from the block $\bar{K}'=(\bar{k_1}',\bar{k_2}',\dots)$ of pseudo random number. 
So the data bit is mapped into one of 2$M$ coherent states randomly.
Quantum state sequences emitted from the transmitter can be described as follows:
\begin{equation}
|\Psi \rangle =|\alpha^j \rangle_1 |\alpha^k \rangle_2 
|\alpha^l \rangle_3 \dots 
\end{equation}
where $|\alpha^j \rangle$ is one of 2$M$ coherent states, $\alpha^j ={\alpha^j}_c +i {\alpha^j}_s$, and $j,k,l\in {\cal{M}}=(1 \sim 2M)$. 
In the phase modulation scheme (PSK), the coherent states are described by positions on a circle in the phase space representation. The radius corresponds to the amplitude or the average photon number per pulse at the transmitter. The positions on the circle correspond to phase information of the light wave. If the number of basis is $M$, then the signal distance between  neighboring states is about 
\begin{equation}
\Delta_{PM} \cong \frac{2\pi |\alpha|}{2M}
\end{equation}
On the other hand, for amplitude  modulation scheme (ASK), it is given by 
\begin{equation}
\Delta_{AM} =\frac{|\alpha_{max}-\alpha_{min}|}{2M}
\end{equation}
For direct intensity modulation of laser diode, 
\begin{equation}
\Delta_{IM} =\frac{|\alpha_{max}|^2-|\alpha_{min}|^2}{2M}
\end{equation}
where $\alpha_{max}$ and $\alpha_{min}$ are the maximum and the minimum amplitude, respectively.
The quantum noise of a coherent state is described by two-dimensional Gaussian distribution  of variance:$1/2$ each, or a one-dimensional Gaussian of variance:$1/4$, if one heterodynes or homodynes respectively. Alice and Bob will design  the number of basis which satisfies 
\begin{equation}
P_e(i, i+1) = \frac{1}{2} - \frac{1}{\sqrt{2\pi}}\int_{0}^{t_0}\exp(-t^2/2) dt
=0.2 \sim 0.5
\end{equation}
where $t_0= \Delta_{PM}/2=\frac{\pi |\alpha|}{2M}$ for the phase modulation scheme, and $t_0= \Delta_{AM}/2$ for amplitude or intensity modulation scheme. This corresponds to the error probability between neighboring states.
But it is not real error probability of Eve. 
The real error probability of Eve depends on her strategy and quantum measurement scheme. 
It is easy to show the numerical examples of the error probability between neighboring states with respect to  signal power and number of basis in amplitude or intensity modulation scheme. For example, $P_e(i, i+1)$ is about 0.45 when  $|\alpha_{max}|=100$, $|\alpha_{min}|=80$, and $M=100$.

\section{Known/chosen plaintext attack}
  A general framework of  attacks on data and on key in the stream cipher is explained  in the reference [19], in which  the probability distribution of the plaintext message in  ciphertext only attacks is uniform, according to the attacker,  and it is non-uniform in  known plaintext attacks.
 If Eve knows the deterministic plaintext, it corresponds to the chosen plaintext attack.
We are concerned with the security when Eve can insert  her deterministic plaintext. So in this section, our discussions will be devoted to the chosen  plaintext attack which is the special case of  known plaintext attacks.

\subsection{Heterodyne measurement}
Let us consider the stream cipher by the general PRNG of the unicity distance $u_{Gn}=f(|K|)$, where $f(|K|)$ is a function of $|K|$. The conventional stream cipher has the following structure.
\begin{equation}
c_i=x_i \oplus {k_i}'
\end{equation}
where ${k_i}'$ is running key.
In this case, if Eve knows some plaintexts ($x_i\in \hat X$ ) and corresponding ciphertext bit sequence,  the output sequence as the ciphertext  corresponds to the running key.
If Eve knows the plaintext bits of $|\hat X|=f(|K|)$ and corresponding ciphertext bits,  she can determine the secret key by the well-known algorithm, and can decode plaintext of the remained bits sequence:
\begin{equation}
|K'|-f(|K|),
\end{equation}
When the PRNG is a LFSR, $f(|K|)=2|K|-1$, and $|K'| \sim 2^{|K|}$ . Since, in general, the $f(|K|)$ is finite in the conventional stream ciphers, they are not secure.

In the case of the quantum stream cipher, the quantum noise effect is unavoidable, because the signal structure in Y-00 protocol to any kind of measurement of Eve has the signal set of non-orthogonal states. In addition, the running key is the decimal number. So Eve cannot get exact ciphertext from her measurement.
Here we assume that Eve can get all energy of light wave from the transmitter at the point close to Alice, and she employs the heterodyne measurement.
By the heterodyne measurement, Eve measures the quadrature amplitudes $\alpha_c$ and $\alpha_s$ putting known plaintext  to decide which basis was used. Since the quadrature amplitudes are non-commutative, the heterodyne measurement corresponds to the simultaneous measurement of non-commuting observables. So the quantum noise is described by the variance $1/2$. Let us assume that the the signal power is large, and the number of basis is large enough.  At that time, even if the measured number for the basis is 5, the true number can be 3 or 7.  On the other hand, when we employ the overlapping selection keying(OSK) as the modulation randomization[7], the errors of measured data of Eve are induced mainly for the neighboring quantum states. That is, even if the measured number for the basis is 5, the true number can be 4 or 6.
These are the minimum requirement for our design. So the error of the basis will be $J > 3$, where $J$ is the number of error basis.
Since the quantum error for each measurement is statistically independent,
 in the individual measurement, the number of combination for $f(|K|)/\log_2 M$ slots which correspond to 
$f(|K|)$ bits in the output of the PRNG is given by
\begin{equation}
Q\cong J^{f(|K|)/\log_2 M}.
\end{equation}
After the measurement, Eve has to transform the decimal number with error into the bit sequence, and start the Berlekamp-Massey algorithm or several known algorithms. Since Eve's data contain error by quantum noise, she cannot get the true secret key from the calculation by the algorithm based only on one trial of the known/chosen plaintext attack. So Eve cannot decode the remained ciphertext sequence.

Here, if Eve can make $Q$ copies of the output sequence of coherent states by cloning procedure, she can try the brute force attack on $Q$ copies, comparing the known/chosen plaintext and each decoded data[7,20]. The near optimum cloning for the coherent state sequence is the beam splitter scheme. It means that Eve has to copy the sequence by means of division of the output light from Alice by $Q$ beam splitters. So the copies are described as follows:
\begin{eqnarray}
|\Psi \rangle &=&|\frac{\alpha^j}{Q} \rangle |\frac{\alpha^k}{Q} \rangle |\frac{\alpha^l}{Q} \rangle \dots \nonumber \\
|\Psi \rangle &=&|\frac{\alpha^j}{Q} \rangle |\frac{\alpha^k}{Q} \rangle |\frac{\alpha^l}{Q} \rangle \dots \nonumber \\
|\Psi \rangle &=&|\frac{\alpha^j}{Q} \rangle |\frac{\alpha^k}{Q} \rangle |\frac{\alpha^l}{Q} \rangle \dots \\
\vdots \nonumber
\end{eqnarray}
Here, when we assume that $\alpha= 100 \sim 1000$, $M=100 \sim 2000$,  $|K|=100 \sim 1000$, the amplitude of the coherent state of Eve becomes $\alpha/Q\sim 0$. That is, the signal to noise ratio is zero. So Eve cannot get any information by the measurement or apply  known/chosen plaintext attacks. 

On the other hand, let us consider that Eve knows the plaintexts of more than $Z$ bits defined by the following equation:
\begin{equation}
Z\equiv \frac{f(|K|)}{\log_2 M}Q.
\end{equation}
In this case, the number of bits of the known/chosen plaintext, which Eve needs, increases exponentially with respect to the key length.
Let $|K'|$ be the output bit length of the PRNG, and let us assume that the communication is stopped at a period of the PRNG. 
If the number of the bits $Z$ is 
\begin{equation}
Z << |K'|/\log_2 M,
\end{equation}
Eve may find the true key by $Q$-times measurements using possible $Q$ keys and input-output data, because the number of keys is reduced to be $Q$ by the first measurement for $f(|K|)/\log_2 M$ bit. 
As an example, when the PRNG is the LFSR, 
\begin{equation}
Z\cong \frac{(2|K|-1)}{\log_2 M}J^{(2|K|-1)/\log_2 M} < 2^{|K|}-1.
\end{equation}
As a result, Eve can find the key at least by the brute force search.
So the original scheme of Y-00 is the exponentially-search-based security against  known/chosen plaintext attacks, when the power of the transmitter is large. 
However, if a PRNG provides
\begin{equation}
Z > |K'|/\log_2 M,
\end{equation}
the brute force attack cannot be completed in a period of the PRNG.
As a result, the success probability of the attack with exponential number of the known/chosen plaintext does not become the unity. We shall here show an example. If we employ the non-linear feedback shift register (NFSR) such as "de Brujin sequence" as the pseudo random generator[10], there exists a sequence with $f(|K|)>> 2|K|$ and the period $2^{|K|}$. In this case, we get
\begin{equation}
Z >> 2^{|K|}/\log_2 M,
\end{equation}
We should emphasize that the security of the PRBG is not essential for the security of Y-00. We only need the large linear complexity, because the security of Y-00 is guaranteed by preventing the trial of the brute force attack itself.

On the other hand, even if we employ the LFSR, it is easy to provide the relation of the Eq(24) by additional randomizations[6,7] such as the breaking of phase locking or the rotation of the axis of the phase plain in the phase modulation scheme, and the sifting of the center line of the amplitude in the intensity modulation scheme.  As a result, we have
\begin{equation}
Q\cong M^{2|K|/\log_2 M}=2^{2|K|}.
\end{equation}
So again the brute force attack cannot be completed  in a period of the PRNG. 
Thus, if the attack is only the brute force attack,  it is at least secure during about $|K'|/\log_2 M$ data bits, even if Eve has the infinite power of computing and the infinite memory capacity. This means that the quantum stream cipher with {\it an appropriate  design} is  secure against the known/chosen plaintext attack, if Eve can only  carry out the brute force attack under the heterodyne measurement.

 On the reuse of the key, for LFSR with $|K|=100$,  the legitimate users  need not to change the key for more than $10^{12}$ years, when the bit rate of the modulator is 1 Gbps. So the quantum stream cipher has no problem with the repetition of the secret key whenever the PRNG is not reset.

\subsection{Indirect measurement}
There is a possibility of the  attack  based on indirect measurements and  post processing  which can reduce the quantum noise effect. 
In fact, there have been some criticisms against Y-00 based on such an indirect measurement[21], but they are wrong.
Here we analyze the subjects related to those criticisms. The typical method of the reduction of noise effects is to measure indirect observable of  the signal, which  a certain modulation scheme connects. Indeed, the $M$-ary keying is taken to be
\begin{equation}
l_i=x_i\oplus \tilde{k}_i
\end{equation}
where $l_i$ is one of two regions separated by an appropriate axis on the phase space or on the line of the strength of the amplitude. 
If the fundamental axis is horizontal, $l_0$ is upper plain, $l_1$ is down plain, $x_i$ is data bit. $\tilde{k}_i$ is 0 for even number and 1 for odd number in the running key of $M$-ary assignment [7,17]. 
For example, 
$(l_i=up, \tilde{k}_i=even) \rightarrow x=1$,  
$(up, odd) \rightarrow x=0$, 
$(down, even) \rightarrow x=0$, 
$(down, odd )\rightarrow x=1$. 
However, we should denote that $\tilde{k}_i$ is the result of the mapping from the running key of decimal number as follows:
\begin{eqnarray}
\bar{k_i}'&=& 1,3,5,\dots \rightarrow \tilde{k}_i=1 \nonumber \\
\bar{k_i}'&=& 2,4,6,\dots \rightarrow \tilde{k}_i=0 
\end{eqnarray}
The essential point of the attack is to measure indirect observable $l_i$. However, since the observable does not contain the information of the data bit, Eve has to try the brute force attack on data in the ciphertext only attack. Moreover, the error of the measurement for $l_i$ is unavoidable. That is, the density operators of signal sets for up and down measurement are 
\begin{eqnarray}
\rho_{up}&=&\sum_{up} \frac{1}{M}|\alpha_{i}\rangle \langle \alpha_{j}|, \\
\rho_{down}&=&\sum_{down} \frac{1}{M}|\alpha_{j}\rangle \langle \alpha_{j}|
\end{eqnarray}
It is easy to show the quantum  limit, which is the most rigorous lower bound of error probability for this signal[12,22]. When the coherent state is mesoscopic $<n> \sim 10000$ and one thousand of $M$ in the phase modulation scheme, the error is about 0.1 percent. As a result, we have $H(X|Y) > H(K)$. If we employ an additional randomization, the error becomes $1/2$.

In the case of the known/chosen plaintext attack, even if the system is error-free, then the measurement results of $l_i$  only tell whether the running key is even or odd. Eve cannot get the true running key from the information of whether it is even or odd. Besides, the system is not error-free. So it does not work. The above results will deny several criticisms based on the noise reduction by the indirect measurement method. 
The detailed discussion is also given in the reference [23].

\subsection{Quantum unambiguous measurement}
Let us discuss the chosen plaintext attack based on a collective quantum unambiguous measurement. Again, Eve knows  plaintexts of $2|K|-1$ bits, and she prepares the quantum unambiguous measurement which can be applied to quantum state sequences of $(2^{|K|}-1)/\log M$.  
When one of quantum state sequences of the set is transmitted from Alice, Eve will measure it by her unambiguous measurement. The success probability is evaluated by an exact calculation and also the following property[8].\\
\\
{\bf Remark}. 
{\it The upper bound of average success probability in the quantum unambiguous measurement is smaller than the  quantum optimum solution in the quantum detection theory for the same state ensemble.}\\
\\
The quantum unambiguous measurement (QUM) for $M$ symmetric coherent states is formulated by  A.Chefles and S.M.Barnett[24]. 
The success probability is given by the following formula.
\begin{equation}
P_D(QUM)= M \min_{k=1,2,3, \dots, M} |c_k|^2
\end{equation}
where
\begin{equation}
|c_k|^2= \frac{1}{M}\sum^{M}_{j=1}e^{2\pi ijk/M}e^{|\alpha|^2(e^{2\pi ij/M}-1)}
\end{equation}
In fact, in the case of the individual measurement, the success probability of the quantum unambiguous measurement on $M$=2000 symmetric coherent states with $(|\alpha|^2=<n>=10000)$ is given by S.van Enk [25] as follows:
\begin{eqnarray}
P_D(QUM)&=&3{\rm x}10^{-12} <\frac{1}{M}=5{\rm x}10^{-4} \nonumber \\
&<& P_D(Bayes)\sim 2{\rm x}10^{-1}
\end{eqnarray}
where $1/M$ is a pure guessing.
And also, the success probability for collective QUM is given by
\begin{equation}
{\bf P_D}(QUM)\sim 0 < 2^{-|K|} < {\bf P_D}(Bayes)
\end{equation}
Thus, the success probability is less than that of pure guessing.
So in general it does not work.

\section{Communication distance}
Here, we analyze how long we can communicate under such a secure condition. Let us assume that the amplitude attenuation parameter of channel between Alice and Bob is $\kappa$. The amplitude of Bob is given by $\kappa \alpha$. When the situation is as follows:
\begin{equation}
\kappa \alpha > \frac{\alpha}{Q},
\end{equation}
even if Eve has a correct key, the error is greater than that of Bob. The signal distance for Bob in the case of phase modulation is given by 
\begin{equation}
d_P=2\kappa|\alpha|
\end{equation}
and, that for the intensity modulation is 
\begin{equation}
d_I=\frac{1}{2}\kappa^2(|\alpha_{max}|^2 - |\alpha_{min}|^2)
\end{equation}
If there is no device noise, the error probability of Bob is given by $d_P$ or $d_I$. The attenuation parameter $\kappa$, which can keep as follows:
\begin{equation}
 {P_e}^B < {P_e}^E,
\end{equation}
determines the communication distance. As a result, the scheme of the intensity modulation can communicate within at least 100 km with 1 Gbps. The application of the optical amplifier will be reported in the subsequent paper. See also the reference [19].

\section{Experimental result}
Several demonstrations of the quantum stream cipher by phase modulation have been reported by the Northwestern University group[17,18,19].
We would like to realize the quantum stream cipher by the intensity modulation[7,8] which is widely used in the conventional optical fiber network system. In order to verify the excellent feature of Y-00 by the intensity modulation scheme, we show  experiment which was done by the Panasonic and our group announced at March 30, 2005.  

Let us give again a brief explanation of the scheme.
The maximum and minimum amplitudes of the transmitter are fixed.
We divide it into 2$M$. So we have $M$ sets of basis state
$\{(A_1,A_2),(B_1,B_2), \dots \}$. 
The total set of basis states is shown in Fig.1. 
Here, the output intensities are square of each amplitudes. 
In addition, we employ OSK[7] which means that  data bits are sent by  switching randomly each basis: $(A_1=0,A_2=1)$ or $ (A_1=1,A_2=0)$, 
$(B_1=0,B_2=1)$ or $ (B_1=1,B_2=0)$, and so on.
The system consists of the distributed-feedback laser diode of the wavelength $\lambda =1.550 \mu m$ and photo diode which works under the 10 Gbps and the room temperature. The linearity  of the laser diode can be applicable  to the analog modulation and the number of basis :$M$ are controlled from 100 to 200. The output power of the laser diode is 0 dBm at the continuous operation, and the launch power is kept between  about -25 $\sim$ -20 dBm by the attenuator. 
The running key is generated by the hardware LFSR with the secret key of 20 bits. The data rate for the modulation is 1 Gbps, and the transmission line is about 20 km spool of single mode fiber.
The decision levels of the decoding system of Bob are automatically controlled by the hardware LFSR with the same secret key as the transmitter. In addition, the systems of three parties are completely synchronized. The difference is only with key or without key. As a result, the detection scheme of Bob is  binary, and that of Eve is $M$-ary.

In this experiment, we assumed that the technology level of Bob and Eve would be the same one. Figure 2 shows the eye patterns for encrypted 1 Gbps of Bob (top) who knows the key and of Eve (bottom) who does not know the key, respectively. This scheme corresponds to ciphertext only attack for Eve.
In these experiments, Bob is located at the end of the 20 Km-long line and Eve is located at the transmitter.
This figure clearly shows that Bob's error performance is better than that of Eve.

More sophisticated experiments with the demonstration of the heterodyne attack and the application to High dense TV system will be reported in the subsequent paper.

\section{Conclusion}
In this paper, we have analyzed some security problems  of the quantum stream cipher by Y-00 protocol, in which the brute force complexity-based security is applied and a certain condition for the design is given to guarantee the security.  As a result, the quantum stream cipher  can be secure against known/chosen plaintext attacks by the heterodyne measurement  or the quantum unambiguous measurement  during $|K'|/log_2M$ of data bits which corresponds to a period of PRNG.
Experimentally, we have implemented and demonstrated the system of the quantum stream cipher by the intensity modulation scheme  with the data rate of 1 Gbps for 20 km long fiber line. However, the demonstrated system has the security in which the security of the conventional stream cipher is  enhanced by quantum noise randomization.
This provides the randomized stream cipher which has high-rate and high-security that  cannot be realized by any kind of the conventional symmetric key cipher. In the subsequent experiments, we will implement several randomizations.

\section*{Acknowledgment}
We are grateful to S.J.van Enk and T.Usuda for discussion, and  to S.Furusawa and T.Ikushima  of Panasonic  for experimental collaboration.  This work was supported by Panasonic funding.

\begin{figure}[htb]
\centerline{
\includegraphics[width=8cm]{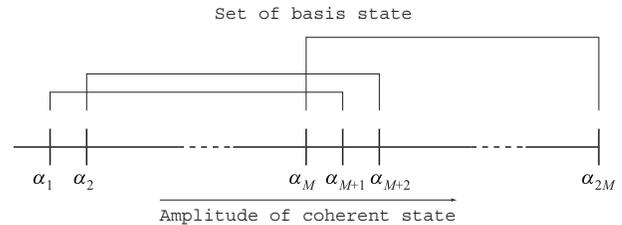}
}
\caption{
\label{fig1}
Design of the basis for  amplitude (intensity) modulation
}
\end{figure}
\begin{figure}[htb]
\centerline{
\includegraphics[width=8cm]{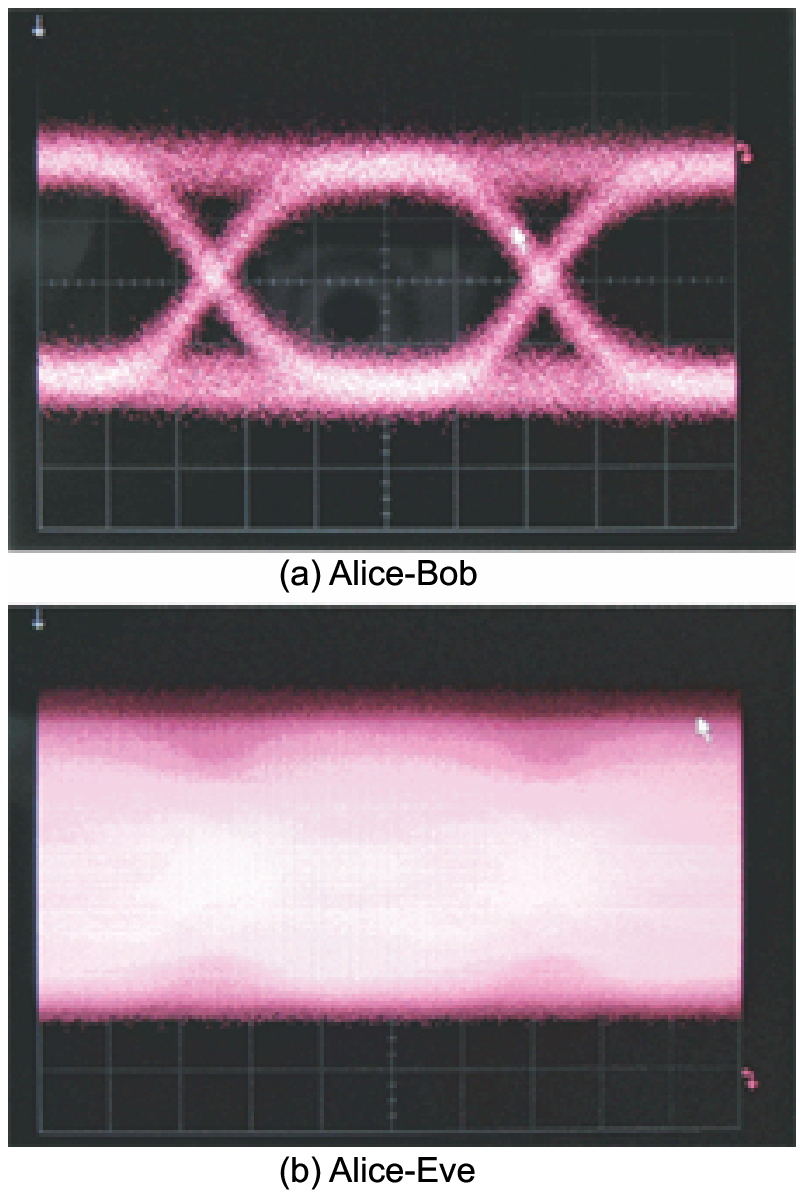}
}
\caption{\label{fig2} (Color Online) 
Eye patterns of Bob(top) and Eve(bottom). The eye-pattern of Eve has no eye-opening, which makes it impossible to discriminate with all of the threshold for the M-ary signals.
}
\end{figure}
\end{document}